\documentclass[twocolumn,superscriptaddress,amsmath,aps]{revtex4}

\usepackage{amssymb,mathrsfs,bm,color,times}
\usepackage{graphicx,color}
\usepackage{bm}
\usepackage[bookmarks=false]{hyperref}
\usepackage{amsmath}
\usepackage{subfigure}
\hypersetup{colorlinks=true,citecolor=blue,linkcolor=blue,urlcolor=blue,pdfstartview=FitH,bookmarksopen=true}

\newcommand{\be}{\begin{equation}}
\newcommand{\ee}{\end{equation}}
\newcommand{\bea}{\begin{eqnarray}}
\newcommand{\eea}{\end{eqnarray}}
\newcommand{\bsube}{\begin{subequations}}
\newcommand{\esube}{\end{subequations}}

\newcommand{\Eq}[1]{Eq.\,(\ref{#1})}

\newcommand{\la}{\langle}
\newcommand{\ra}{\rangle}

\newcommand{\beq}{\begin{equation}}
\newcommand{\eeq}{\end{equation}}
\newcommand{\beqn}{\begin{eqnarray}}
\newcommand{\eeqn}{\end{eqnarray}}

\newcommand{\bsub}{\begin{subequations}}
\newcommand{\esub}{\end{subequations}}

\begin{document}

\title{Postselected amplification and photon recycling
applied to optical sensing of magnetic fields}

\author{Yazhi Niu}
\affiliation{Center for Joint Quantum Studies and Department of Physics,
School of Science, \\ Tianjin University, Tianjin 300072, China}

\author{Jialin Li}
\affiliation{Research Center for Quantum Physics and Technologies,
Inner Mongolia University, Hohhot 010021, China}
\affiliation{School of Physical Science and Technology,
Inner Mongolia University, Hohhot 010021, China}

\author{Lupei Qin }
\email{qinlupei@tju.edu.cn}
\affiliation{Center for Joint Quantum Studies and Department of Physics,
School of Science, \\ Tianjin University, Tianjin 300072, China}

\author{Xin-Qi Li}
\email{xinqi.li@imu.edu.cn}
\affiliation{Research Center for Quantum Physics and Technologies,
Inner Mongolia University, Hohhot 010021, China}
\affiliation{School of Physical Science and Technology,
Inner Mongolia University, Hohhot 010021, China}


\date{\today}

\begin{abstract}
We apply the combined technique of postselected amplification and photon-recycling
to an optical setup of magnetic field precision measurement.
We propose two recycling schemes
and carry out analytic expressions for
the amplified signal and measurement sensitivity.
The results show significant improvement of performance over conventional measurement.
The underlying reason is twofold.
On one aspect,
introducing the technique of recycling
eliminates the shortcoming of data discarding in postselection,
thus maintains similar noise level of conventional measurement (without postselection).
On the other aspect, performing intentional postselection
within the recycling framework, which was originally proposed
in the context of gravitational wave detection, can amplify the signal.
Thus, the measurement signal-to-noise ratio is enhanced.
\end{abstract}

\maketitle

\section{Introduction}

In the field of precision metrology, detection of weak magnetic fields
plays important role in diverse fields of fundamental science
and practical technologies \cite{Bud07,Bud17,PXH21a,PXH21b,Ger20,Fot20,Bow22,Kaz20}.
In a recent work \cite{Niu24}, we proposed to embed the full optical
atomic magnetometer into an optical Mach-Zehnder interferometer (MZI).
It was shown that the Faraday rotation (FR) angle of the probe laser light
can be largely amplified, by means of postselecting
the path-information state of the laser photons when passing through the MZI.
In the presence of photo-detector saturation and/or polarization cross talk
in the polarizing-beam-splitter performance, the amplified FR angle
is expected to make the postselection measurement (PSM) scheme
outperform the conventional measurement (without postselection).

The basic idea of the PSM scheme is letting a small change
in a system-meter coupling parameter be converted into
a large change in a meter variable \cite{AAV1,DJ14,How14,Llo20,Shu22,Yang23}.
This ``large change" effect comes at the price
of measuring only a small postselected fraction of the events,
e.g., photons, in the optical interferometry case.
As a result of this gain and loss balance, the PSM scheme has
the same signal-to-noise ratio (SNR) of the measured parameter,
despite that it can perform better than the conventional measurement
in the presence of certain technical limitations,
such as beam jitter noise or detector saturation, as demonstrated by experiments
in the ultrasensitive optical beam displacement and deflection measurements
\cite{Kwi08,How09a,How10a,How10b,Lun17,ZLJ20}.

In order to further improve the measurement sensitivity (i.e., enhance the SNR),
the recycling technique,
which was initially proposed in laser-interferometric
gravitational-wave detectors \cite{Dre83,Meer88,Meer91,Dan97}
and has been successfully implemented in practice \cite{Abb09,Aba11},
was borrowed and demonstrated in the postselected weak measurements
\cite{Kwiat13,Kwiat15,Kwiat21,Guo16}.
After involving the technique of recycling,
the photons that would have previously exited the interferometer
through the bright port, are now repeatedly recycled back
to the input port and reenter the interferometer.
As a result, the number of postselected photons exiting the dark port,
each carrying large Fisher information, is considerably increased
in comparison with the standard PSM, leading thus to a much better SNR.

In this work, beyond the study in Ref.\ \cite{Niu24},
we consider applying this combined technique,
say, the recycled PSM (rPSM) scheme,
to the optical FR-based precision measurement of weak magnetic fields.
Actually, there existed two types of recycling scheme.
One is the continuous-wave power recycling
\cite{Dre83,Meer88,Meer91,Dan97,Kwiat15,Guo16},
which, in the context of PSM,
relies on designing destructive interference to trap the photons
that would have previously exited the interferometer
through the bright port inside the interferometer,
so that almost the entire input light exits the dark port of the interferometer.
Another is the pulsed recycling \cite{Kwiat13,Kwiat21},
which relies on sending a sequence of laser pulses
and using Pockels cell (PC) and polarization optics
to trap each pulse in the interferometer.
In this work, we focus on applying the pulsed recycling scheme
to postselected measurement of weak magnetic fields,
while remaining the application of the continuous-wave
power recycling technique as study in a separate future work.

The paper is organized as follows.
In Sec.\ II, we outline the theory of PSM without recycling.
Then, we propose and carry out analytic results
for two pulsed recycling schemes in Sec.\ III and IV.
In Sec.\ V, we present some numerical results and,
in particular, the results of finite number of recycling times.
Finally, we summarize the work with some discussions in Sec.\ VI.

\section{Postselected Amplification without Recycling}

To be specific, and for the convenience of principle demonstration,
rather than the nowadays state-of-the-art atomic magnetometer \cite{Bud07,Bud17,PXH21a,PXH21b},
we follow the setup studied in Ref.\ \cite{Yin21},
to consider a Faraday crystal (FC) placed between two electric coils,
as shown in Fig.\ 1(a), which can introduce a relative phase
between the left circularly polarized light and the right one,
or equivalently,
cause a Faraday rotation (FR) of a linearly polarized light,
with the FR angle proportional to the magnetic field to be measured.
In this work, as in Ref.\ \cite{Niu24}, we propose to embed
this magnetometer into the MZI, as shown in Fig.\ 1(b),
for the purpose of postselected amplification of the FR angle,
by employing the path states of the MZI to realize postselection.

\begin{figure}
\includegraphics[scale=0.85]{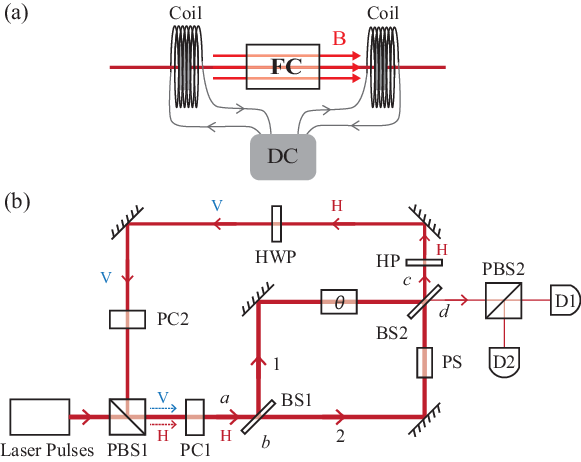}
  \caption{
(a)
Schematic of an optical magnetometer, in which a Faraday crystal (FC) is placed
between two electric coils, which can introduce a relative phase
between the left circularly polarized light and the right one,
equivalently, cause a Faraday rotation (FR) of a linearly polarized light,
with the FR angle proportional to the magnetic field to be measured.
(b)
Following Ref.\ \cite{Niu24}, the magnetometer (characterized by the FR angle $\theta$)
is proposed to embed into an MZI. However, beyond Ref.\ \cite{Niu24},
an external optical circuit is introduced
to recycle the discarding light of postselection in the bright port
back to the input port, and re-inject it into the MZI.
Optical elements in this set-up:
BS (beam splitter); PBS (polarizing beam splitter);
PC (Pockels cell); HP (horizontal polarizer);
HWP (half wave plate); D1 and D2 (two photo-detectors).
Cooperative action of these elements can fulfill the requirement of the pulse-recycling scheme.
In particular, in order to properly control the polarization of light,
while the PC2 in the external circuit is keeping on ``off" state,
the PC1 in the input port should be switched from the initial ``off" state
(to inject the initial pulse with ``H" polarization into the interferometer),
to the later ``on" state
(to ensure the successive recycling of light re-injected with ``H" polarization).
In the final stage of recycling, the PC2 is switched on
to convert the polarization of light from ``V" to ``H",
releasing the residual light in the interferometer
via transmitting through the PBS1,
and allowing next pulse's input and recycling.       }
\end{figure}

Let us denote the two entrance path states
of the MZI as $|a\ra$ and $|b\ra$,
the two path states inside the MZI as $|1\ra$ and $|2\ra$,
and the two exit port path states as $|c\ra$ and $|d\ra$.
Also, the polarization states of light are denoted as $|H\ra$ and $|V\ra$.
Inside the MZI, in path ``1", the polarization of the light experiences a FR,
owing to the Faraday crystal and magnetic field, which can be described by
acting $\hat{U}_{1}= e^{i\theta |1\ra\la 1| \hat{\sigma}_{x}}$,
on the joint state of path and polarization.
Here, $\hat{\sigma}_{x}= |V\ra\la H|+{\rm h.c.}$, and $\theta=VBl$.
$B$ is the magnetic field to be measured, $l$ is the length of the Faraday crystal,
and $V$ is the Verdet constant \cite{Yin21}.
In path ``2", a phase shift is introduced, intentionally,
for the purpose of postselection, which is described
by the state evolution operator $\hat{U}_{2}= e^{i\beta|2\ra\la 2|}$.

Let us consider a laser pulse fed into the MZI through
the first beam splitter (BS1),
and denote the initial state as $|\psi\ra=|a\ra|H\ra$.
One may notice that the laser pulse is usually
described as an optical coherent state $|\alpha\ra$,
which is a quantum superposition of different photon-number states
and has an average photon number $N=|\alpha|^2$.
In our analysis, the relevant degrees of freedom of the light
are the path and polarization of the photons,
thus we only introduce the states of the relevant degrees of freedom.
The laser pulse transmits through the MZI and is postselected
into the exit port ``d" (a dark port) for further measurement.
The state evolution and postselection are described as
\bea\label{D-1}
|\tilde{\psi}_{d1}\ra
= \hat{M}_{d}\hat{S}_{2}\hat{U}_{2}\hat{U}_{1}\hat{S}_{1}|\psi\ra  \,.
\eea
Here, the role of the BS1 is described by the scattering operator
$\hat{S}_{1}=\frac{1}{\sqrt{2}}\{ (|1\ra\la a|+|2\ra\la a|)
+(|1\ra\la b|-|2\ra\la b|)\}$.
Similarly, for the second beam splitter (BS2),
the scattering operator is given by
$\hat{S}_{2}=\frac{1}{\sqrt{2}}\{ (|c\ra\la 1|+|d\ra\la 1|)
+(|c\ra\la 2|-|d\ra\la 2|)\}$.
The postselection into exit port ``d"
is described by the measurement operator $\hat{M}_{d}=|d\ra\la d|$.
After simple algebras, we obtain
\bea\label{D-1-b}
|\tilde{\psi}_{d1}\ra = |d \ra \hat{M}_{-} |H\ra
\equiv  |d \ra |\tilde{\phi}_{d1}\ra   \,.
\eea
Here, we introduced
$\hat{M}_{-}=(e^{i\theta\hat{\sigma}_{x}}-e^{i\beta})/2$.
The success probability of postselection is given by
\bea
p_{d1}=   \la\tilde{\phi}_{d1}|\tilde{\phi}_{d1}\ra
=(1-\cos\theta\cos\beta)/2 \,.
\eea
The normalized state is 
$|{\phi}_{d1}\ra=|\tilde{\phi}_{d1}\ra /\sqrt{p_{d1}}$.
In this state, the probability of finding the ``V" component can be obtained through
$P_{V}={\rm Tr}(|V\ra\la V| \rho_{d1})$,
while $\rho_{d1}= |{\phi}_{d1}\ra \la{\phi}_{d1}|$.
Explicitly, we obtain
\bea
P_{V}=\sin^{2}\theta/(4p_{d1})\equiv \sin^{2}{\tilde{\theta}} \,.
\eea
In this context, by analogy with conventional measurement,
we defined an effective angle $\tilde{\theta}$,
which may be termed as postselection-amplified FR angle.
This result indicates that, with the decrease of $p_{d1}$
(by modulating the postselection parameter $\beta$),
$\tilde{\theta}$ can be much larger than the true FR angle $\theta$,
realizing thus a remarkable amplification effect.
For smaller $\theta$, the amplification effect can be more prominent,
for instance, it can be amplified by several orders of magnitude \cite{Niu24}.
Physically speaking, the reason of this amplification is that
each postselected photon, in the state  $|{\phi}_{d1}\ra$,
contains more quantum Fisher information
than the photon before postselection, 
owing to an essential role of quantum interference. 
This result also reflects the key feature of
the recently advocated postselection filtering technique,
which has been termed as postselected metrology \cite{Llo20,Shu22,Yang23}.  

Moreover, we carry out the sensitivity of the PSM scheme as follows,
for quantum-limited measurement (i.e., neglecting various practical imperfections).
Actually, the above result, $P_{V}=\sin^{2}{\tilde{\theta}}$,
is the average of the projection operator
$\hat{P}_V=|V\ra\la V|$ in the state $\rho_{d1}$,
say, $P_{V}=\la \hat{P}_V\ra = {\rm Tr}(\hat{P}_V \rho_{d1})$.
Similarly, we can obtain $\la \hat{P}^2_V\ra$
and evaluate the quantum variance
$\delta P_V = \sqrt{\la\hat{P}_V^2\ra -(\la\hat{P}_V\ra)^2 }$.
On the other hand, from $\sin\tilde{\theta}= \sqrt{P_V}$,
we have $\delta \tilde{\theta}
=(2\sqrt{P_V} \cos\tilde{\theta} )^{-1} \delta P_V$.
Then, after simple algebras, we obtain $\delta \tilde{\theta}=1/2$.
Further, consider $N$ incident photons
(thus $\widetilde{N}_1=p_{d1}N$ postselected photons).
Based on the probability $P_V$
measured using the $\widetilde{N}_1$ postselected photons,
the estimate uncertainty of $\tilde{\theta}$
is $\delta \tilde{\theta}=(2\sqrt{p_{d1} N})^{-1}$.
Noting that $\tilde{\theta}\simeq \theta/(2\sqrt{p_{d1}})$
(in the case of small $\theta$ and $\tilde{\theta}$),
the sensitivity is obtained as
$\delta \tilde{\theta} / \tilde{\theta}=(\theta \sqrt{N})^{-1}$,
which is the same of conventional measurement (without postselection).  

\section{Recycling Scheme (I)}

The above PSM scheme suffers a problem
that a large number of photons are discarded into port ``c".
The dramatic point is that,
the fewer photons postselected into port ``d" can contain almost 
all the information of all photons before postselection.
Actually, this feature promises some advantages
in the presence of technical limitations \cite{Niu24}.
In the following, beyond Ref.\ \cite{Niu24},
we consider to recycle the discarded photons in port ``c",
by guiding them back to the input port ``a",
and re-injecting them into the interferometer.
As the first scheme of recycling, referred to as {\it scheme (I)},
we design an external recycling circuit as sketched in Fig.\ 1(b).

The 1st-round trip has been described in Sec.\ II.
For the 2nd-round trip, we first determine the state of the photon
exited from port ``c" and passed through the horizontal polarizer (HP).
It is given by
$|\tilde{\psi}_{c1}\ra
= \hat{M}_H \hat{M}_{c}\hat{S}_{2}\hat{U}_{2}\hat{U}_{1}\hat{S}_{1}|\psi\ra$,
where $\hat{M}_H=|H\ra\la H|$ and $\hat{M}_c=|c\ra\la c|$
are two measurement operators. More explicitly, we obtain
\bea\label{c-1}
|\tilde{\psi}_{c1}\ra = |c \ra M_{+} |H\ra
\equiv  |c \ra |\tilde{\phi}_{c1}\ra  \,,
\eea
with ${M}_{+}=(\cos\theta+e^{i\beta})/2$.
The probability of obtaining this state is
$p_{c1}=\la \tilde{\phi}_{c1}|\tilde{\phi}_{c1}\ra
=(1+2\cos\theta\cos\beta+\cos^{2}\theta)/4$.
Assuming that the recycled photon suffers certain loss 
in the external optical path (with a ratio denoted by $L$), 
and has a delayed phase $e^{i\epsilon}$
(compared to the 1st-round injected photon),
we obtain the state of the 2nd-round postselected photon in port ``d"
as $|\tilde{\psi}_{d2}\ra = |d \ra |\tilde{\phi}_{d2}\ra$,
while $|\tilde{\phi}_{d2}\ra$ reads as
\bea
|\tilde{\phi}_{d2}\ra
=\hat{M}_{-}(\sqrt{1-L}\,e^{i\epsilon}{M}_{+})|H\ra  \,.
\eea
Repeating the same considerations, the state of
the $j_{\rm th}$-round photon postselected in port ``d" is given by
\bea\label{d-j}
|\tilde{\phi}_{dj}\ra
=\hat{M}_{-}(\sqrt{1-L}e^{i\epsilon}{M}_{+})^{j-1}|H\ra  \,.
\eea
Straightforwardly, the probability of obtaining this state is
$p_{dj}=\la \tilde{\phi}_{dj}|\tilde{\phi}_{dj}\ra =\{(1-L)p_{c1}\}^{j-1}p_{d1}$,
while the normalized state is
$|\phi_{dj}\ra=|\tilde{\phi}_{dj}\ra/\sqrt{p_{dj}}$.

Formally, the maximum number $n_{\rm max}$ of recycling can be determined by the condition
$(\sqrt{1-L}e^{i\epsilon}{M}_{+})^{n_{\rm max}}=0$ (actually $n_{\rm max}\to\infty$).
Then, the total probability that the photon is postselected
into port ``d" (via $n_{\rm max}$ times of recycling) is
\bea\label{Pd-I}
P_{d}=\sum\limits_{j = 1}^{{n_{\rm max}}} {{p_{dj}}}=\frac{p_{d1}}{1-(1-L)p_{c1}} \,.
\eea

In the recycling process,
the optical element HP will cause photon loss,
owing to filtering only the ``H" polarization.
In the first round trip, the loss probability is
$\gamma_{1}=\frac{1}{4}\sin^{2}\theta$.
In general, the $j_{\rm th}$-round loss caused by the HP is
$\gamma_j = [(1-L) p_{c1}]^{j-1} \gamma_1$.
Summing all these losses, we obtain
\bea\label{Loss-S}
\Gamma=\frac{\gamma_1}{1-(1-L) p_{c1}} \,.
\eea
In the case $L=0$ (no extra ``external" loss),
one can check that, $P_{d}+\Gamma=1$, as it should be,
by noting that $p_{d1} + p_{c1} + \gamma_1 =1$.

In the above summation of deriving the total probability $P_d$,
the following considerations have been involved.
Quantum mechanically, the multiple outgoing states (the train of the recycled
exiting wavepackets) in port ``d" should be in quantum superposition,
i.e., described by the (unnormalized) state
$|\tilde{\phi}_{d}\ra = \sum^{n_{\rm max}}_{j=1} |\tilde{\phi}_{dj}\ra $.
The reason is that, before {\it considering measurement} by the classical photodetectors,
all the optical elements in the MZI and the recycling setup are unitary transformation devices.
The component state $|\tilde{\phi}_{dj}\ra$ in the summation
corresponds to the $j_{\rm th}$ exiting pulse (wavepacket) of the photon.
The different wavepackets are separated in space, before being measured by photodetectors.
Then, the quantum measurement will collapse the state $|\tilde{\phi}_{d}\ra$
onto an individual component state, e.g., $|\tilde{\phi}_{dj}\ra$.
As a result, the state, after accounting for the effect of measurement,
is a mixed state given by
\bea\label{d}
\rho_{d}
=\sum\limits_{j = 1}^{{n_{\max }}}\frac{p_{dj}}{P_{d}}
|\phi _{dj}\ra \la \phi _{dj}|
\,.
\eea
Based on this state, one can obtain
the probability of finding the ``V" polarization
through computing $P_V = \mathrm{Tr}(\hat{P}_V \rho_{d})=\la\hat{P}_V \ra$.
Explicitly, we obtain
\bea\label{PV}
P_V =\frac{\sin^{2}\theta}{4\{1-(1-L)p_{c1}\}P_{d}}   \,.
\eea
Inserting the result of $P_{d}$, given by \Eq{Pd-I},
into this expression, we obtain
$P_V=\sin^2\theta/(4 p_{d1}) \equiv \sin^{2}\tilde{\theta}$,
which is the same as the result without recycling.
In both cases, the signal (the FR angle) can be considerably amplified,
say, from $\theta$ to $\tilde{\theta}$.

However, the recycling procedure re-collects photons
exiting from port ``c" and re-injects them into the MZI.
Since each photon emerging out from port ``d" carries
rich Fisher information about $\theta$,
the entire postselected photons in port ``d" in the presence of recycling
should contain much more Fisher information than the case without recycling,
and the sensitivity of measurement is expected to be greatly improved.
Indeed, we can carry out the sensitivity as follows.
First, we obtain
$\delta P_{V}=\sqrt{\la \hat{P}^{2}_{V}\ra - \la{{\hat{P}_V}} \ra^{2}}
=\sin\tilde{\theta}\cos\tilde{\theta}$.
Second, from $\sin\tilde{\theta}=\sqrt{P_{V}}$, we obtain
$\delta \tilde{\theta}=(2\sqrt{P_{V}}\cos\tilde{\theta})^{-1}\delta P_{V}=1/2$.
Further, consider $N$ photons in the laser pulse, which is actually
a coherent state $|\alpha\ra$ (thus $N=|\alpha|^2$).
The postselected photon number in the presence of recycling is $\widetilde{N}=P_{d}N$.
Therefore, after accounting for the shot noise of photons,
we have $\delta\tilde{\theta}=(2\sqrt{P_d N})^{-1}$.
Finally, the sensitivity of measurement is obtained as
\bea\label{sens-I}
\frac{\delta\tilde{\theta}}{\tilde{\theta}}
= \frac{[1-(1-L)p_{c1}]^{1/2}}{\sqrt{N}\theta} \,.
\eea
Compared with
$\delta\tilde{\theta} / \tilde{\theta} = 1/(\sqrt{N}\theta)$,
i.e., the sensitivity of conventional measurement
or the PSM without recycling,
the result of \Eq{sens-I} has the same scaling with $N$.
However, the extra factor in the numerator can be very small.
For instance, consider $L=0$ (no extra loss in the recycling circuit).
This factor is simplified to $\sqrt{\gamma_1 + p_{d1}}$.
Both $\gamma_1$ and $p_{d1}$ can be very small.
This simply means a high sensitivity of the rPSM scheme.
The basic reason for this sensitivity enhancement is twofold.
On one hand, the signal is amplified, from $\theta$ to $\tilde{\theta}$.
On the other hand, the uncertainty $\delta\tilde{\theta}=(2\sqrt{P_d N})^{-1}$
is comparable to conventional measurement (without postselection),
which is much smaller than the result $\delta\tilde{\theta}=(2\sqrt{p_{d1}N})^{-1}$
of the PSM without recycling.

\section{Recycling Scheme (II)}

\begin{figure}
  \includegraphics[scale=0.8]{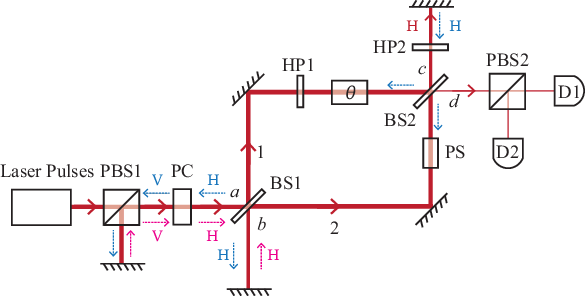}
  \caption{
Internal recycling scheme, without introducing the external circuit as shown in Fig.\ 1(b).
The role of HP1 and HP2 is filtering the ``V" component of the light,
allowing only the ``H" component of light transmitted through them.
The role of the PC in the input port, first set in ``off" state
to allow the transmission of the initial pulse with ``H" polarization,
and later switched to ``on" state, is converting the polarization
from ``H" to ``V" and from ``V" to ``H",
in order to resend the outgoing light back into the interferometer,
with ``H" polarization.   }
\end{figure}

In this section, we consider an alternative recycling scheme,
referred to as {\it scheme (II)}, as shown in Fig.\ 2.
Starting with the input state in port ``a", $|\psi\ra=|a\ra|H\ra$,
we obtain the 1st-round outgoing states in ports ``d" and ``c" as
$|\tilde{\phi}_{d1}\ra=\hat{M}_{-} |H\ra$
and $|\tilde{\phi}_{c1}\ra= M_{+} |H\ra$,
which are the same as in scheme (I).
In the recycling scheme (II), the photon in port ``c"
is returned back to ports ``a" and ``b",
inversely along the inside paths of the MZI,
resulting in a joint state of path and polarization given by
\bea
|\Psi(\theta)\ra
 =\frac{1}{2}\left[ \cos\theta(|a\ra+|b\ra)+e^{i\beta}(|a\ra-|b\ra)\right] M_{+}|H\ra
 \,.
\eea
Re-sending this state into the MZI, the 2nd-round outgoing state in port ``d"
is obtained through $|\tilde{\psi}_{d2}\ra
= \hat{M}_{d}\hat{S}_{2}\hat{U}_{2}\hat{U}_{1}\hat{S}_{1}|\Psi(\theta)\ra$.
Explicitly, we obtain $|\tilde{\psi}_{d2}\ra = |d\ra|\tilde{\phi}_{d2}\ra$,
while $|\tilde{\phi}_{d2}\ra=\hat{Q}_{-}M_{+}|H\ra$,
with $M_{+}$ given already in scheme (I) and $\hat{Q}_{-}
=\frac{1 }{2}(\cos\theta e^{i\theta\hat{\sigma}_{x}}-e^{i2\beta})$.
Further, for latter use, we introduce
$D_{-}=\la H|\hat{Q}^{\dagger}_{-}\hat{Q}_{-}|H\ra
=\frac{1}{4}(1+\cos^{2}\theta-2\cos^{2}\theta\cos2\beta)$.
Similarly, the 2nd-round outgoing state in port ``c" and after HP2
is obtained through
$|\tilde{\psi}_{c2}\ra
= \hat{M}_{H} \hat{M}_{c}\hat{S}_{2}\hat{U}_{2}\hat{U}_{1}\hat{S}_{1}|\Psi(\theta)\ra
= |c\ra|\tilde{\phi}_{c2}\ra$,
with $|\tilde{\phi}_{c2}\ra=Q_{+}M_{+}|H\ra$
and $Q_{+}=\frac{1}{2}(\cos^{2}\theta +e^{i2\beta})$.
For latter use, we also introduce $D_{+}=|Q_{+}|^2=\frac{1}{4}
(1+\cos^{4}\theta+2\cos^{2}\theta\cos2\beta)$.

Iteratively, following the recycling procedure,
we obtain the $j_{\rm th}$-round outgoing state in port ``c" as
\bea \label{j-C}
|\tilde{\phi}_{cj}\ra = Q_{+}^{j-1}M_{+}|H\ra  \,,
\eea
and the outgoing state in port ``d" as
\bea \label{j-D}
|\tilde{\phi}_{dj}\ra =
\hat{Q}_{-}{Q}^{j-2}_{+}M_{+}|H\ra  ~~ (j\geq2)  \,.
\eea
We further introduce the normalized state
$|{\phi}_{dj}\ra=\frac{1}{\sqrt{p_{dj}}}|\tilde{\phi}_{dj}\ra$,
in which the normalization factor
is associated with postselection probability
$p_{dj}=\la \tilde{\phi}_{dj}|\tilde{\phi}_{dj}\ra=D_{-}D_{+}^{j-2}p_{c1}$.
Formally,
the maximum recycling number $n_{\rm max}$
is determined through $Q_{+}^{n_{\rm max}-2}M_{+}=0$.
The total probability of collecting photon in port ``d" is
$P_{d}=p_{d1}+\sum_{j = 2}^{n_{\rm max}}p_{dj}$.
Explicitly, we obtain
\bea
P_{d}=p_{d1} + (1-D_{+})^{-1} D_{-}\, p_{c1} \,.
\eea

In this recycling scheme,
the optical elements HP1 and HP2 will cause photon loss.
In the first round, the loss probability caused by HP2 is
$\gamma_{1}=\frac{1}{4}\sin^{2}\theta$.
In the second round, the loss probability caused by HP1 and HP2 is
$\gamma_{2}=\frac{1}{4}\sin^{2}\theta\,(2+\cos^{2}\theta)\,p_{c1}$.
And, in general, the $j_{\rm th}$ round loss caused by HP1 and HP2 is
$\gamma_j = \frac{1}{4}\sin^{2}\theta\,(2+\cos^{2}\theta)D_{+}^{j-2}\,p_{c1}$.
Summing all these losses, we obtain
\bea\label{Loss-S}
\Gamma=\frac{1}{4} \sin^{2}\theta
\left[ 1 + (2+\cos^{2}\theta) \frac{p_{c1}}{1-D_{+}} \right] \,.
\eea
One can check that, $P_{d}+\Gamma=1$, as it should be.

For the same reason as explained in recycling scheme (I),
the multiple outgoing states postselected in port ``d"
constitute a mixture, given by
\bea\label{d}
\rho_{d}
=\sum_{j = 1}^{n_{\rm max}}\frac{p_{dj}}{P_d}
|\phi _{dj}\ra \la \phi _{dj}|   \,.
\eea
Using this state, one can compute the probability
of ``V" polarization measured in port ``d" as
$P_V = \mathrm{Tr}( \hat{P}_V\rho_{d})
= \eta\sin^{2}\theta  \equiv \sin^{2}\tilde{\theta}$, with $\eta$ given by
\bea\label{eta2}
\eta=\frac{1}{4P_d}
\left(1+ \cos^{2}\theta \frac{p_{c1}}{1-D_{+}} \right) \,.
\eea

In order to carry out the result of sensitivity,
we first obtain the variance
$\delta P_{V}=\sqrt{\la \hat{P}^{2}_{V}\ra - \la{{\hat{P}_V}} \ra^{2} }
=\sin\tilde{\theta}\cos\tilde{\theta}$.
Then, from $\sin\tilde{\theta}=\sqrt{P_{V}}$, we obtain
$\delta \tilde{\theta}
=(2\sqrt{P_{V}}\cos\tilde{\theta})^{-1}\delta P_{V} = 1/2 $.
Again, as in scheme (I), we assume $N$ photons in the laser pulse.
The photon number from the recycling-assisted postselection in port ``d" is $P_dN$.
Then, the estimate uncertainty of $\tilde{\theta}$,
after accounting for the shot noise of photons,
is $\delta \tilde{\theta}= (2\sqrt{P_d N})^{-1}$.
Finally, we obtain the sensitivity as
\bea
\frac{\delta\tilde{\theta}}{\tilde{\theta}}
= (\sqrt{(1+x)N} \theta)^{-1} \,.
\eea
Here, we introduced $x=\cos^{2}\theta \, p_{c1} (1-D_{+})^{-1}$.
As to be shown numerically in the following,
compared to the sensitivity of conventional measurement,
the extra factor $1/\sqrt{1+x}$ can be much smaller than 1,
resulting thus in an improved sensitivity of measurement.

\section{Numerical Results and Finite Number of Times of Recycling}

\begin{figure}
  \includegraphics[scale=0.8]{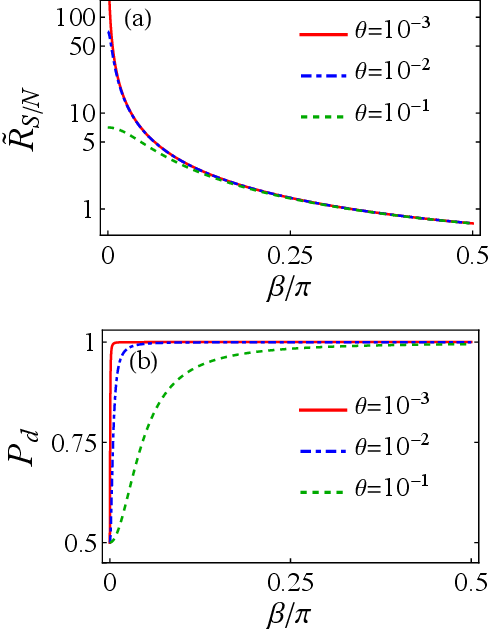}
  \caption{
(a)
Scaled SNR of the rPSM scheme (I), associated with the schematic plot of Fig.\ 1.
It shows that stronger postselection (smaller $\beta$)
can result in a better enhancement of the SNR,
especially for measuring weaker magnetic field,
which corresponds to the smaller FR angle $\theta$.
This result indicates the remarkable role of postselection
in the recycling measurements.
(b)
Total postselection probability of photons into the dark exit port ``d",
after accounting for all the infinite number of times of recycling.
The larger loss of $P_d$, for larger $\theta$, is because of
the stronger loss of photons caused by the polarization filtering element HP,
in the bright exit port ``c".   }
\end{figure}

\begin{figure}
  \includegraphics[scale=0.8]{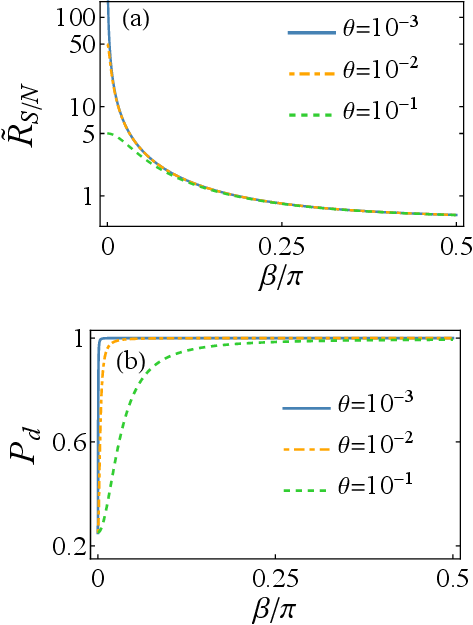}
  \caption{
(a)
Scaled SNR of the rPSM scheme (II), associated with Fig.\ 2;
and (b),
the total postselection probability of photons into the dark exit port ``d",
after accounting for all the infinite number of times of recycling.
The qualitative behaviors of both are similar to
the results shown in Fig.\ 3, for the rPSM scheme (I).   }
\end{figure}

To see more clearly the effect of postselection in the presence of photon recycling,
we plot the result of the enhanced factor of the signal-to-noise ratio (SNR),
$\widetilde{R}_{S/N}=R_{S/N}/R^{\rm cm}_{S/N}$, in Fig.\ 3 for scheme (I).
Here, the SNR is defined as the reciprocal of the sensitivity, i.e.,
$\tilde{\theta}/\delta\tilde{\theta}$.
Importantly, compared to the recycling technique involved in the laser interferometer
of gravitational wave detection \cite{Dre83,Meer88,Meer91,Dan97},
the {\it intentional postselection}, i.e., postselecting photons into a dark port
with amplified Fisher information about the parameter under estimation,
can further enhance the SNR, even greatly in principle.
The enhancement effect is more evident for smaller $\theta$
and stronger postselection (with smaller $\beta$), as shown in Fig.\ 3(a).
The basic reason is twofold.
On one aspect, the signal is greatly amplified,
from $\theta$ to $\tilde{\theta} = \eta\theta$,
with $\eta=1/(4p_{d1})$ being very large
if the postselection probability $p_{d1}$ is small.
On the other aspect,
the total postselection probability $P_d$,
as numerically shown in Fig.\ 3(b),
can make the postselected photon number
$\widetilde{N}=P_d N$ comparable to the input photon number
(much larger than $\widetilde{N}_1=p_{d1} N$,
i.e., the postselected photon number without recycling);
thus the shot noise is maintained at
the same level of the conventional measurement, while the signal is amplified.

Similarly, in Fig.\ 4,
we show the results of the enhanced factor of SNR $\widetilde{R}_{S/N}$
and the total postselection probability $P_d$, for the recycling scheme (II).
Despite that the specific expressions of these quantities
are different from their counterparts of scheme (I),
qualitative behaviors are essentially the same.

In practice, in the pulsed recycling scheme, not a single pulse,
but a sequence of laser pulses are fed into the interferometer.
Then, before the arriving (and entering the interferometer) of next pulse,
the previous one has only been recycled a finite number of times through the interferometer.
In the following, we present results of 
finite number of times of recycling, for a single pulse.

Let us denote the interval time between two nearest pulses by $T$,
and a round trip time of each recycling process by $T_r$,
which is determined by the physical size of the setup.
Following Ref.\ \cite{Kwiat13}, we assume $T=1$ ms and $T_r=10$ ns.
Then the number of recycling times of each pulse is $n=T/T_r=10^5$.
We will show that this number is large enough,
which can result in negligible residual light in the interferometer.
Moreover, in scheme (I), one can release the residual light by switching on the PC2
(just before the arrival of next laser pulse),
to let the ``H"-polarized residual light transmit through the PBS1 (but not reenter the MZI).

Following the line of derivations outlined in Sec.\ II, III and IV,
one can obtain the results
for finite number of times of recycling, quite straightforwardly.
For scheme (I), instead of the result of \Eq{PV},
for infinite number of times of recycling,
we obtain the result of $n$ times of recycling as
\bea\label{PVn-1}
P^{(n)}_V
=\frac{(1-y^n) \sin^{2}\theta}{4(1-y) P^{(n)}_{d}}   \,.
\eea
Here, we introduced $y=(1-L)p_{c1}$,
and obtained the accumulated total probability
\bea
P^{(n)}_{d}=p_{d1}(1-y^n)/(1-y) \,.
\eea
Actually, we can define
$P^{(n)}_V=\eta^{(n)} \sin^2\theta \equiv \sin^2\tilde{\theta}^{(n)}$
and find that the signal amplification factor $\eta^{(n)}=1/(4p_{d1})$
is the same as previous result of scheme (I), being independent of $n$.
We can also carry out the corresponding SNR.
Noting that the total number of postselected photons in port ``d"
is $\widetilde{N}^{(n)}=N P^{(n)}_{d}$,
the estimate uncertainty of $\tilde{\theta}^{(n)}$ is then obtained as
$\delta \tilde{\theta}^{(n)} = (2\sqrt{\widetilde{N}^{(n)}})^{-1}$.
Accordingly, we obtain the SNR as
$R^{(n)}_{S/N}=2\theta\sqrt{\widetilde{N}^{(n)}\eta^{(n)}}$,
and the enhanced factor as
$\widetilde{R}^{(n)}_{S/N}= R^{(n)}_{S/N}/R^{\rm cm}_{S/N} = \sqrt{P^{(n)}_{d}\eta^{(n)}}$.

For scheme (II), instead of \Eq{PVn-1},
we obtain the result of finite number of recycling as
$P^{(n)}_V=\eta^{(n)} \sin^2\theta \equiv \sin^2\tilde{\theta}^{(n)}$,
with the amplification factor given by
\bea
\eta^{(n)} = \frac{1}{4} \left( \frac{\kappa_n + \cos^2\theta\, p_{c1}}
{\kappa_n p_{d1} + D_-\, p_{c1}} \right)  \,.
\eea
Here we introduced $\kappa_n=(1-D_+)/(1-D^{(n-1)}_+)$. Notice that,
being different from scheme (I), the amplification factor of scheme (II)
depends on $n$ (the number of recycling times).
We also obtain the result of $P_d^{(n)}$,
the total postselected probability of $n$ times of recycling, as
\bea
P_d^{(n)} = p_{d1}+p_{c1}D_-/\kappa_n \,.
\eea
Further, for the corresponding SNR, we obtain the result of
$R^{(n)}_{S/N}=\theta\sqrt{(1+\chi_n)N}$,
while $\chi_n=\cos^2\theta\, p_{c1}/\kappa_n$.
Then, the enhanced factor of SNR can be simply expressed as
$\widetilde{R}^{(n)}_{S/N} = R^{(n)}_{S/N}/R^{\rm cm}_{S/N}
= \sqrt{1+\chi_n} / 2 = \sqrt{P^{(n)}_{d}\eta^{(n)}} $.

\begin{figure}
  \includegraphics[scale=0.85]{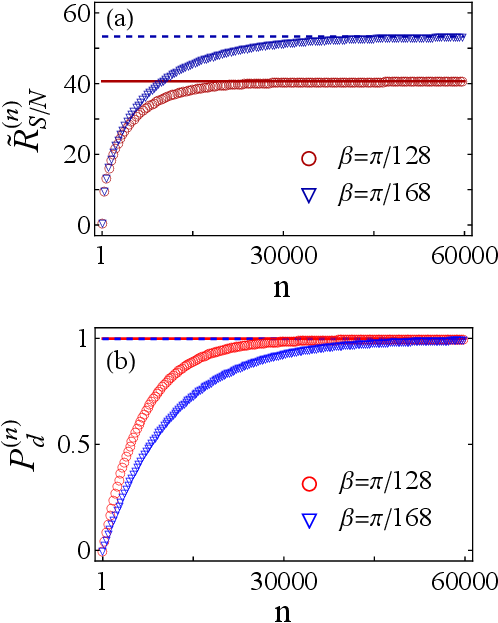}
  \caption{
(a) Scaled SNR $\widetilde{R}^{(n)}_{S/N}$
and (b) the total postselection probability $P^{(n)}_d$,
for finite ($n$) times of recycling based on the rPSM scheme (I).
Results of two postselection parameters are displayed and compared with
the results of infinite number of times of recycling (the solid and dashed lines),
showing that both can become the same for sufficiently large $n$.
In the main text, the number of recycling times $n=10^5$
is estimated based on certain practical parameters.   }
\end{figure}

\begin{figure}
  \includegraphics[scale=0.85]{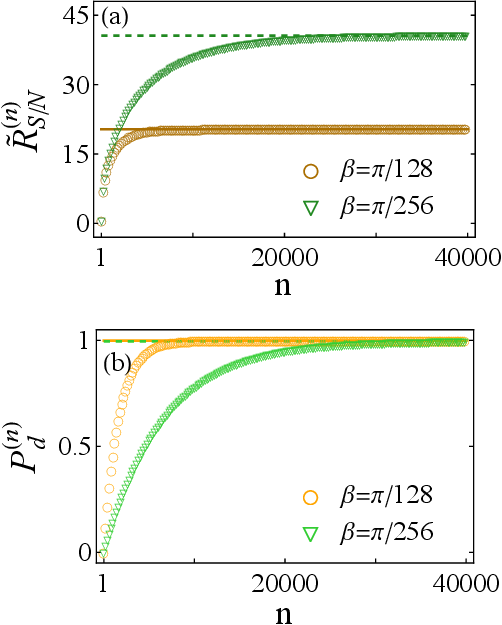}
  \caption{
(a) Scaled SNR $\widetilde{R}^{(n)}_{S/N}$
and (b) the total postselection probability $P^{(n)}_d$,
for finite ($n$) times of recycling based on the rPSM scheme (II).
Other explanations are the same as in the caption of Fig.\ 5.   }
\end{figure}

In Figs.\ 5 and 6, for both recycling schemes,
we plot the results of $\widetilde{R}^{(n)}_{S/N}$ and $P_d^{(n)}$
against the number of recycling times, and compare with the results
of infinite number of times of recycling (the solid and dashed lines),
We find that, with the increase of $n$,
the results will converge fast to the limits of infinite times of recycling.
As estimated based on certain practical parameters,
the number of recycling times of $n=10^5$
can be achieved between two pulses,
which results thus in negligibly weak residual light in the interferometer.
Moreover, in scheme (I),
one can further release the residual light by switching on the PC2
(just before the arrival of next laser pulse),
to let the ``H"-polarized residual light transmit through the PBS1 (but not reenter the MZI).
In scheme (II), we are unable to design such release protocol.
However, from the simulated results shown in Fig.\ 6,
we expect that, after the $n=10^5$ times of recycling of the laser pulse,
the residual light is negligibly weak and does not affect the work of the next pulse.

\section{Summary and Discussion}

We have presented a study on applying the combined technique 
of postselected amplification and photon-recycling
to an optical setup of magnetic field precision measurement.
Based on the optical Mach-Zehnder interferometer,
we proposed two schemes of pulse recycling 
and carried out analytic expressions for
the amplified signal and measurement sensitivity.
The results show significant improvement of signal-to-noise ratio
over both conventional measurement (without postselection)
and pure postselected measurement (without recycling).

In practice, in order to control the polarization of light in the input port,
the PC1 in the recycling scheme (I) (see Fig.\ 1(b))
and the PC in the recycling scheme (II) (see Fig.\ 2)
should be switched from the initial ``off" to the later ``on" state,
for properly completing the successive processes of recycling.
For this purpose, their switching time should be shorter than
the round trip time (e.g., $T_r=10$ ns) of the light pulse.
Actually, the manufacturer-quoted switching time
of the Pockels cell is about 3 ns \cite{Bed21},
which satisfies thus the requirement mentioned here.
About the PC2 in the recycling scheme (I)
in the external circuit, its final step switching on
during the recycling of each laser pulse is less restricted,
since this operation just converts the polarization of light from ``V" to ``H",
letting thus the residual light in the interferometer
be released (via transmitting through the PBS1).
After this, one can start next pulse's input and recycling.

The pulse-recycling schemes analyzed in present work
are most suitable for measuring static magnetic field.
In this case, each recycling trip of the pulse
suffers the same FR angle $\theta$ when transmitting through the Faraday crystal.
For the case of time dependent magnetic field,
if the field varies slowly (compared to the recycling time),
the pulse-recycling schemes might still work,
but may require more careful and more complicated treatment (a job for future study).
A better way to solve this problem might be designing
	 a continuous wave power-recycling scheme
\cite{Dre83,Meer88,Meer91,Dan97,Kwiat15,Guo16}, 
which works
in the steady state of the continuous laser light in the presence of recycling.
The steady state is actually a quantum superposition
of the large number of recycled components of light.
The design and analysis of this alternative recycling scheme is also left for our future study.

Finally, we remark that, despite that the power-recycling technique
was initially proposed and later implemented
in gravitational wave detection \cite{Dre83,Meer88,Meer91,Dan97,Abb09,Aba11},
introducing postselection strategy to amplify signal
together with the recycling technique
is seemingly not yet considered in this field,
and in other relativistic gravity effects detection \cite{Ra17,Sab17,Sab18}.

\vspace{0.5cm}
{\flushleft\it Acknowledgements.}---
This work was supported by the Startup Project of Inner Mongolia University
and Inner Mongolia Autonomous Region (No.\ 10000-23112101/299),
and the NNSF of China (No.\  11974011).

\vspace{0.5cm}

\end{document}